\documentclass[usenatbib]{mn2e}
\usepackage{graphics}

\begin{document}

\title[Near-IR Properties of I-Drops in the HUDF]
{Near-Infrared Properties of I-Drop Galaxies in the 
Hubble Ultra Deep Field}
\author[E.~R.~Stanway et al.]{Elizabeth R. Stanway\,$^{1}$, 
Richard G. McMahon\,$^{1}$ \& Andrew J. Bunker\,$^{1,2}$ \\
$^{1}$\,Institute of Astrophysics, University of Cambridge,
Madingley Road, Cambridge, CB3\,0HA, U.K.\\
{\tt email:ers@ast.cam.ac.uk, rgm@ast.cam.ac.uk}\\
$^{2}$\,School of Physics, University of Exeter, Stocker Road, Exeter, EX4 4QL,
 U.K.\\
{\tt email:bunker@astro.ex.ac.uk}}

\date{Submitted to MNRAS; version 1.1}

\maketitle

\begin{abstract}

We analyse near-infrared {\em HST}/NICMOS $F110W (J)$ and $F160W (H)$ band
photometry of a sample of 27 $i'$-drop candidate $z\simeq6$ galaxies
in the central region of the HST/ACS Ultra Deep Field (HUDF).  The
infrared colours of the 20 objects not affected by near neighbours
are consistent with a high redshift interpretation. This suggests that
the low redshift contamination of this $i'$-drop sample is smaller
than that observed at brighter magnitudes where values of 10-40\% have
been reported.  The $J-H$ colours are consistent with a slope flat in
$f_\nu$ ($f_\lambda\propto\lambda^{-2}$), as would be expected for an
unreddened starburst.  There is, however, evidence for a marginally
bluer spectral slope ($f_\lambda\propto\lambda^{-2.2}$) which is
perhaps indicative of an extremely young starburst ($\sim10$ Myr old)
or a top heavy initial mass function and little dust.
 The low levels of contamination, median photometric redshift of
$z\sim6.0$ and blue spectral slope, inferred using the near-infrared
data, supports the validity of the assumptions in our earlier
work in estimating the star formation rates
and, that the majority of the i-drop candidates galaxies lie at $z\sim6$.

\end{abstract}

\section{Introduction}
\label{sec:intro}

In recent years the observational horizon has expanded rapidly and
radically for those observing distant galaxies.  Large format
red-sensitive detectors on wide field imaging instruments, the new
generation of 8m class telescopes and the refurbished Hubble Space
Telescope ({\em HST}), have pushed the limits to which we can
routinely detect star-forming distant galaxies progressively from
redshifts of one to beyond $z=6$.  At the highest redshifts currently
accessible, narrow band emission lines searches using the
Lyman-$\alpha$ line have moved on from redshifts of 4
\citep{hm96,fu03}, to 5.7 \citep{hu99} and now reach to $z\sim6.5$
\citep{ko03,hu02}.  Photometric redshifts
\citep[e.g.][]{lan96,co97,fe99} are now routinely used to process
large datasets and identify high redshift candidates. An application
of this method, the continuum based Lyman-break photometric technique
pioneered at $z\sim3$ by \citep{guh90,st95}, has been extended
progressively to z$\sim4$ and $z\sim5$ \citep[e.g.][]{st99,br04},
spectroscopically confirmed, and using $i'$-band drop selection
further extended to $z\sim6$
\citep{st03,st04a,st04b,bo03a,bb03,bo04,di04,yan03}.

The `classical' Lyman break technique used by Steidel et al.\ (1996)
uses three filters, one redward of rest frame Lyman-$\alpha$
($\lambda_\mathrm{rest}>1216$\AA), a second in the spectral region
between rest-frame Lyman-$\alpha$ and the rest-frame Lyman limit
(912\AA) and a third at $\lambda_\mathrm{rest}<912$\AA. At z$\sim$3 the
technique relies on the ubiquitous step in the spectra of the stellar
component of galaxies at 912\AA\ due to photospheric absorption,
supplemented by optically-thick Lyman limit system absorption caused
by neutral hydrogen in the galaxy in question or in the intervening
IGM.  At higher redshifts the evolution in the Lyman-$\alpha$ forest
absorption, particularly in the spectral region 912-1216\AA\ means
that the effective break migrates redward to the Lyman-$\alpha$
region.  Recently this search technique has been extended to $z\sim6$
by various authors and $i'$-drop samples have been used to constrain
the star formation history of the universe (as derived from rest-frame
UV luminous, starbursting stellar populations, e.g. Stanway, Bunker \&
McMahon 2003; Giavalisco et al.\ 2004; Bouwens et al.\ 2004; Bunker et
al.\ 2004).

Any such estimation, however, is dependent on the 
characteristic properties of the $i'$-drop population. Although
Steidel and coworkers \citep{st99,st01} have successfully shown that a
Lyman-break technique cleanly selects galaxies at intermediate
redshift ($z\approx3-4$), the $i'$-drop varient of the technique lacks
a second colour to constrain the redshift distribution and is subject
to contamination from both low redshift elliptical galaxies and cool
dwarf stars.  Until recently, these factors have been essentially
unconstrained by observational data.

The {\em HST}/ACS Ultra Deep Field (HUDF, Beckwith, Somerville and Stiavelli
2004), imaging an 11 arcmin$^2$ region of the sky to faint magnitudes
in the $F435W(B)$, $F606W(v)$, $F775W(i')$ and $F850LP(z')$ bands, has
now allowed a clean sample of $i'$-drop objects to be defined and
constraints to be placed upon its luminosity function \citep{b04}. The
{\em HST}/NICMOS treasury programme, complimentary to the HUDF, has
imaged the central region of the ACS field to faint magnitudes at
wavelengths of 1.1 and 1.6 microns, allowing the infrared properties of
the $i'$-drop population (in particular their rest-frame ultra-violet
spectral slope, faint-end contamination and luminosity at 1500\AA) 
 to be determined. In this paper we discuss the infrared
properties of the $i'$-drop population defined in \citet{b04}, and
their implications for the nature of these objects.

We adopt a $\Lambda$-dominated, `concordance' cosmology with
$\Omega_{\Lambda}=0.7$, $\Omega_{M}=0.3$ and $H_{0}=70\,h_{70} {\rm
km\,s}^{-1}\,{\rm Mpc}^{-1}$. All magnitudes in this paper are quoted
in the AB system \citep{og83} and the \citet{ma95} prescription,
extended to $z=7$, is used where necessary to estimate absorption due
to the intergalactic medium. Beyond $z=7$ absorption is assumed
to decrease the transmitted flux by 98\% \citep[the decrement observed in
the flux of $z>6$ quasars identified by the Sloan Digital Sky Survey, ][]{fa03}
.

\section{HST Observations and Catalog Construction}
\label{sec:obs}

\subsection{{\em HST}/NICMOS Observations}
\label{sec:nicmos}

\begin{figure}
\resizebox{0.48\textwidth}{!}{\includegraphics{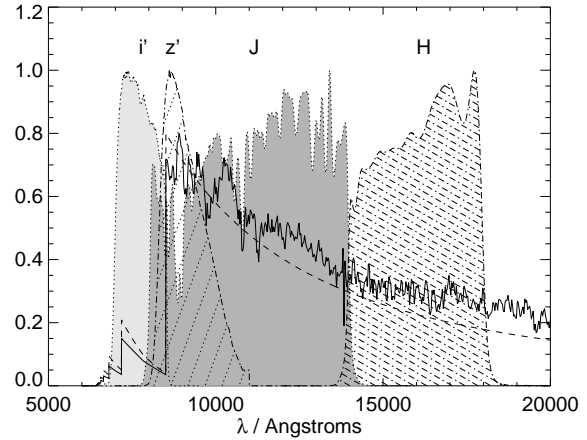}}
\caption{Normalised transmission profiles of the $F775W(i')$ and
$F850LP(z')$ filters on the ACS instrument on HST and the $F110W(J)$
and $F160W(H)$ filters of NICMOS. Note the significant overlap between
the $J$ band and both lower wavelength filters. Overplotted on these
are the profiles of a starburst galaxy placed at $z\simeq6$ (solid
line, Kinney et al.\ 1996) and also a power-law profile with
$f_\lambda\propto\lambda^{-2.0}$ (dashed line, see section
\ref{sec:zbeta}). The \citet{ma95} prescription for intergalactic
absorption at $z=6$ has been applied to both spectra.}
\label{fig:filters}
\end{figure}

In this paper we utilise the publically released, pipeline-reduced
images from the NICMOS UDF treasury programme\footnote{Programme
  GO-9803; PI Thompson; 144 orbit allocation; Available from {\tt
    http://www.stsci.edu/hst/udf}}.  These were taken with the NICMOS
3 camera on {\em HST}, in the $F110W(`J')$ and $F160W(`H')$ filters, with
exposures of 8 orbits per band per pointing giving an average exposure
time of 21500 seconds per pixel in each image.  The transmission
profiles of these filters are illustrated in figure \ref{fig:filters}
and differ from those of standard ground-based $J$ and $H$ filters,
tuned to atmospheric transmission windows. In particular the $F110W$
filter is wide, extending down to 8000\AA.

NICMOS 3 uses a $256\times256$ pixel detector with a scale of
0\farcs20 per pixel, giving a field of view of
$51\arcsec\times51\arcsec$. The images discussed in this paper have
been distortion corrected and drizzled onto a grid of 0\farcs09 per
pixel.  The available NICMOS fields cover $2.3\times2.3$ arcmin giving
a total area of 5.2 arcmin$^2$ (45\% of that of the {\em HST}/ACS HUDF) to a
depth of $J_{AB}=27.73 (3\sigma)$ and $H_{AB}=27.48 (3\sigma)$ in a
2\arcsec\ diameter aperture. They comprise mosaics of $3\times3$
individual NICMOS pointings and the edge regions (with low exposure
times) have been trimmed.  Further details on the use for NICMOS for
deep surveys can be found in \citet{th99} which describes the results
of 49 orbit $J$ and $H$ band observations of a single NICMOS pointing
of the HDF-N \citep{wi96}.

The near-IR ($J$ and $H$ band) magnitudes of objects in these images
are calculated using the photometric calibration provided by the
NICMOS UDF team, with photometric zeropoints given by $J_{AB}=23.4034$
and $H_{AB}=23.2146$\footnote{Note that the photometric calibration of
  the NICMOS instrument appears to have changed significantly from
  that reported in earlier work (zeropoints of $J_{AB}$= 22.7 and
  $H_{AB}$=22.8), probably due to the installation of a new cryostat.
  The older zeropoints have remained in use in the on-the-fly
  calibration software of the MAST archive. AB magnitudes may be
  converted to approximate Vega magnitudes using the prescription:
  $J_{AB}=J_{\mathrm{Vega}}+0.98$, $H_{AB}=H_{\mathrm{Vega}}+1.31$
  \citep{ko83}}. The point spread function of unresolved objects
(measured on the images) were 0\farcs4 FWHM (4 drizzled pixels) in the
$J$-band and 0\farcs45 FWHM (5 drizzled pixels) in $H$.

\subsection{Catalog Construction}
\label{sec:cat}

Catalog construction from the imaging data was performed utilising
v2.3.2 of the SExtractor photometry software package
\citep{ba96}. For object identification, we demanded at least 5 adjacent
pixels above a flux threshold of $2\sigma$ per pixel. Catalogs were
trained in the $J$ band and Sextractor was then used in two-image mode
to evaluate the equivalent magnitude of each object in the $H$ band.

Objects in the resulting catalog were then matched to those in a
second catalog of sources detected in the $z'$ ACS HUDF image, with a
match between two objects being assigned if the astrometry agreed to
better than 0\farcs27 (3 pixels on the drizzled $J$ band image, one
undrizzled NICMOS pixel).  The agreement between astrometric systems
was generally better than 0\farcs1 although the discrepancy increases
with distance from the field centre.  

Given the depth of the HUDF images, the narrower point spread function
and smaller pixel scale ACS (PSF and plate scale are both 0\farcs05
for ACS compared to a plate scale of 0\farcs2 and a PSF of 0\farcs4
for NICMOS) can lead to difficulties with this process, since objects
identified as single sources on the $J$ and $H$ band images may be
resolveable into several distinct galaxies, or components of a single
galaxy, on the optical imaging of the HUDF.  As a result the $z'$ flux
corresponding to a single $J$-selected `source' may be underestimated.

The SExtractor software performs source
deblending based on a multi-thresholding algorithm as described in
detail in \citet{ba96}.  This process involves rethresholding the flux
of each set of connected pixels above the detection threshold into 30
levels spaced between that threshold and the peak flux. At each level,
the source may be divided into two components if the flux in each
component exceeds a given threshold. As tests in \citet{ba96}
indicate, such an approach may fail if there is no saddlepoint in the
flux profile of the blended system (i.e. for objects seperated by less
than two times the instrumental point spread function, or
approximately 0.8\arcsec\ in this case). To verify this, we performed
simulations in which a compact artificial galaxy of $J=25.0$ was
inserted at varying distances from real objects with $23<J<27$ in the
NICMOS UDF images.  As expected, source magnitudes were successfully
recovered for equal magnitude objects seperated by $\approx$0\farcs8,
and for galaxies seperated by 1\farcs2 even with a two magnitude
difference between real and artificial galaxies. On the other hand,
the software could not deblend objects closer than these seperations.
Hence the photometry of $i'$-drops with brighter neighbours
within 1\arcsec\ (indicated with an asterisk in table \ref{tab:phot})
is not used in the analysis in sections \ref{sec:disc} \&
\ref{sec:sfh}.

We use deblended ``total'' magnitudes, rather than fixed
aperture magnitudes in the $J$ and $H$ bands. 
SExtractor calculates the
``total magnitude'' by computing an elliptical aperture whose
ellipticity and elongation are defined by the second order moments of
the object's isophotal light distribution. The flux is measured in an
elliptical aperture with 2.5 times the elongation calculated
above. Hence the measured flux is independent of an assumed radial
light profile. 

The use of different photometric apertures in different observation
bands may lead to low surface brightness flux being missed in one or
more bands. We attempted to quantify this by comparing the $z$-$J$
colours obtained using ``total'' magnitudes with those obtained using
fixed circular apertures, and corrected for missing flux assuming a
compact radial light profile as appropriate in the $z'$ band.  We
found that the resulting colours agree to within approximately 0.2
magnitudes. Given the consistent colours obtained, we decided to adopt
the ``total'' magnitudes rather than assuming a radial light profile
for this source population. However, we note that without knowledge of
the accurate radial light profile some uncertainty remains in the
fraction of flux emitted below the detection threshold in each band.



\section{Near-IR Properties of $z\simeq6$ Candidate Objects}
\subsection{Near-IR properties of bright I-drops in the HUDF}
\label{sec:Idrop}

In a previous paper \citep{b04}, we reported a catalogue
of 54 $z\simeq6$ galaxy candidates, including
one confirmed $z=5.83$ galaxy (Bunker \#20104, SBM03\#1, SiD002, GLARE
1042 in Bunker et al.\ 2004, Stanway et al.\ 2004a, Dickinson et al.\
2004 and Stanway et al.\ 2004b).  Each $i'$-drop object in this selection
satisfied $(i'-z')_{AB}>1.3$ and $z'_{AB}<28.5$ in the HST/ACS images
of the HUDF field. Of this catalog, 27 objects lie in region surveyed
by the NICMOS UDF (consistent with the 45\% of the ACS HUDF area
covered by this near-IR imaging.)

The near-IR photometry of these high redshift candidates is presented
in table \ref{tab:phot} with $3\sigma$ limits on the magnitudes
indicated where appropriate.  The photometry of seven of the objects
is likely to be affected by bright near neighbour objects at lower
redshift.


Interestingly several multicomponent systems of galaxies are observed,
perhaps implying that we are observing merger-triggered star
formation.  As illustrated by figure \ref{fig:stamps} three $i'$-drop
galaxies, 24228, 23972 and 24123, form a close system (spanning
$\approx1\arcsec$) and hence have not been resolved as individual
objects in the near-IR (where each pair separated by just 3 undrizzled
pixels). As a result, the average colour of this galaxy group, Group
1, has been calculated by summing the optical fluxes of the three
components. We note that the majority of near-IR emission is to the
north of the group centre, suggesting that object 23972 may be fainter
than its neighbours in the near-IR bands.

Similarly, the galaxy pairs 44154 \& 44194 (Group 2) and 42929 \&
42806 (Group 3) are each treated as a single system, their optical
flux summed and compared to the blended near-IR flux of each system.
Hence the reported flux for each of these groups of $i'$-drop galaxies
is a system total rather than that of any particular galaxy.

\begin{table*}
\begin{tabular}{l|c|c|c|c|c|c|c|c}
\hline\hline
ID & RA \& Dec (J2000) & $i'_{\rm AB}$ & $z'_{\mathrm AB}$ & $J_{\mathrm AB}$ & $H_{\mathrm AB}$ & $(z'-J)_{\mathrm AB}$ & $(J-H)_{\mathrm AB}$ & $z_\mathrm{phot}$ \\
\hline\hline
Group 1$^1$  &  03 32 34.29  -27 47 52.8  &   28.06 $\pm$   0.15  &  26.41 $\pm$ 0.05  &  26.23  $\pm$  0.10     &  26.09 $\pm$ 0.11     &  0.18 $\pm$ 0.11 &  0.14 $\pm$ 0.15 & 5.93$^{+0.24}_{-0.26}$ \\
Group 2$^2$  &  03 32 36.46  -27 46 41.5  &   28.72 $\pm$   0.22  &  26.36 $\pm$ 0.05  &  26.66  $\pm$  0.11     &  25.82 $\pm$ 0.07     & -0.33 $\pm$ 0.12 &  0.84 $\pm$ 0.13 & 5.75$^{+0.24}_{-0.12}$ \\
Group 3$^3$  &  03 32 37.46  -27 46 32.8  &   29.74 $\pm$ 0.40    &  27.27 $\pm$ 0.09  &  26.53  $\pm$ 0.13      &  26.29 $\pm$ 0.13     &  0.94 $\pm$ 0.16 &  0.24 $\pm$ 0.18 & 6.65$^{+0.35}_{-0.11}$ \\
   20104     &  03 32 40.01  -27 48 15.0  &   26.99 $\pm$   0.04  &  25.35 $\pm$ 0.02  &  25.54  $\pm$  0.04     &  25.51 $\pm$   0.05   & -0.19 $\pm$ 0.04 &  0.03 $\pm$ 0.06 & 5.82$^{+0.01}_{-0.01}$\\
   23516$^*$ &  03 32 34.55  -27 47 56.0  &   28.57 $\pm$   0.10  &  27.04 $\pm$ 0.05  &  26.82  $\pm$  0.13     &  26.85 $\pm$   0.16   &  0.22 $\pm$ 0.13 & -0.03 $\pm$ 0.21 & 5.73$^{+0.08}_{-0.06}$\\
   25941     &  03 32 33.43  -27 47 44.9  &   29.30 $\pm$   0.18  &  27.32 $\pm$ 0.06  &  27.52  $\pm$  0.16     &  27.24 $\pm$   0.15   & -0.20 $\pm$ 0.17 &  0.28 $\pm$ 0.22 & 5.91$^{+0.09}_{-0.07}$\\
   26091$^*$ &  03 32 41.57  -27 47 44.2  &   29.74 $\pm$   0.25  &  27.38 $\pm$ 0.06  &  26.44  $\pm$  0.12     &  26.33 $\pm$   0.13   &  0.94 $\pm$ 0.13 &  0.11 $\pm$ 0.18 & 6.67$^{+0.15}_{-0.06}$\\
24458$^\dag$ &  03 32 38.28  -27 47 51.3  &   29.11 $\pm$   0.15  &  27.51 $\pm$ 0.07  &      $>$27.7            &      $>$27.4          & $<$0.19          &                  & 5.81$^{+0.09}_{-0.06}$\\
   49117$^*$ &  03 32 38.96  -27 46 00.5  &   29.77 $\pm$   0.26  &  27.74 $\pm$ 0.08  &  26.36  $\pm$  0.10     &  25.35 $\pm$   0.06   &  1.34 $\pm$ 0.13 &  1.01 $\pm$ 0.12 & 6.82$^{+0.05}_{-0.05}$\\
   27270     &  03 32 35.06  -27 47 40.2  &        $>$30.4        &  27.83 $\pm$ 0.08  &  27.32  $\pm$  0.16     &  27.30 $\pm$   0.20   &  0.51 $\pm$ 0.18 &  0.02 $\pm$ 0.26 & 6.34$^{+0.30}_{-0.15}$\\
14751$^\dag$ &  03 32 40.92  -27 48 44.8  &   29.39 $\pm$   0.17  &  27.87 $\pm$ 0.09  &      $>$27.7            &      $>$27.4          & $<$0.14          &                  & 5.78$^{+0.13}_{-0.07}$\\
 35084$^*$   &  03 32 44.70  -27 47 11.6  &   29.86 $\pm$   0.28  &  27.92 $\pm$ 0.09  &      $>$27.7            &      $>$27.4          & $<$0.22          &                  & 5.93$^{+0.15}_{-0.17}$\\
46503$^\dag$ &  03 32 38.55  -27 46 17.5  &   29.43 $\pm$   0.20  &  27.94 $\pm$ 0.09  &      $>$27.7            &      $>$27.4          & $<$0.24          &                  & 5.77$^{+0.15}_{-0.09}$\\
 19953$^*$   &  03 32 40.04  -27 48 14.6  &   29.50 $\pm$   0.21  &  27.97 $\pm$ 0.09  &      $>$27.7            &      $>$27.4          & $<$0.27          &                  & 5.78$^{+0.15}_{-0.09}$\\
21111$^\dag$ &  03 32 42.60  -27 48 08.9  &   29.69 $\pm$   0.24  &  28.02 $\pm$ 0.10  &      $>$27.7            &      $>$27.4          & $<$0.32          &                  & 5.83$^{+0.23}_{-0.11}$\\
46223$^\dag$ &  03 32 39.87  -27 46 19.1  &        $>$30.4        &  28.03 $\pm$ 0.10  &      $>$27.7            &      $>$27.4          & $<$0.33          &                  & 6.08$^{+0.23}_{-0.32}$\\
22138$^\dag$ &  03 32 42.80  -27 48 03.3  &        $>$30.4        &  28.03 $\pm$ 0.10  &      $>$27.7            &      $>$27.4          & $<$0.33          &                  & 6.08$^{+0.23}_{-0.32}$\\
 46234       &  03 32 39.86  -27 46 19.1  &        $>$30.4        &  28.05 $\pm$ 0.10  &      $>$27.7            &      $>$27.4          & $<$0.35          &                  & 6.08$^{+0.24}_{-0.33}$\\
12988$^*$    &  03 32 38.49  -27 48 57.8  &        $>$30.4        &  28.11 $\pm$ 0.11  &      $>$27.7            &      $>$27.4          & $<$0.41          &                  & 6.08$^{+0.25}_{-0.35}$\\
24733$^\dag$ &  03 32 36.62  -27 47 50.0  &        $>$30.4        &  28.15 $\pm$ 0.11  &      $>$27.7            &      $>$27.4          & $<$0.45          &                  & 6.08$^{+0.25}_{-0.36}$\\
21530$^\dag$ &  03 32 35.08  -27 48 06.8  &   30.24 $\pm$   0.39  &  28.21 $\pm$ 0.12  &      $>$27.7            &      $>$27.4          & $<$0.51          &                  & 5.96$^{+0.21}_{-0.32}$\\
35271$^\dag$ &  03 32 38.79  -27 47 10.9  &   29.77 $\pm$   0.26  &  28.44 $\pm$ 0.14  &      $>$27.7            &      $>$27.4          & $<$0.74          &                  & 5.70$^{+0.25}_{-0.13}$\\
 22832$^*$   &  03 32 39.40  -27 47 59.4  &        $>$30.4        &  28.50 $\pm$ 0.15  &      $>$27.7            &      $>$27.4          & $<$0.80          &                  & 6.08$^{+0.34}_{-0.50}$\\
\hline\hline 
\multicolumn{2}{l}{Stack (mean)$^\dag$}   & 30.14 $\pm$ 0.17 & 28.13 $\pm$  0.05 & 28.56 $\pm$   0.16 &    28.45  $\pm$   0.16 & -0.43 $\pm$ 0.17 & 0.11 $\pm$ 0.23 & \\
\end{tabular}\\
\begin{flushleft}

$^*$ Photometry may be affected by near neighbour objects.\\
$^\dag$ Object has contributed to the stack of undetected objects, the mean properties of which are given in the last line.\\
$^1$\,$^2$\,$^3$ $J$ and $H$ band fluxes of these groups of objects are blended (see text for details). Photometry is given for the group as a whole. Group 1 comprises objects 24228, 24123 \& 23972. Group 2 comprises objects 42929 \& 42806. Group 3 comprises objects 44154 \& 44194.\ 
 \end{flushleft}

\caption{Photometry of \citet{b04} candidate objects lying within the
NICMOS UDF Field. Near-IR magnitudes given are total magnitudes. $i'$ and $z'$ magnitudes were
measured in 0\farcs5 apertures and corrected to total magnitudes as
described in \citet{b04}, except for those cases where the $i'$-drop galaxies form small groups.  
In these cases the total magnitude of the group is reported. Upper limits (3$\sigma$, measured in a 2\arcsec\ diameter aperture) are given for the $J$ and $H$ bands where appropriate. ID numbers are those given in \citet{b04}. 
Photometric redshifts were calculated as described in
section \ref{sec:zphot} and the 99\% ($3\sigma$) confidence intervals
are shown.}
\label{tab:phot}
\end{table*}

\begin{figure*}
\resizebox{0.48\textwidth}{!}{\includegraphics{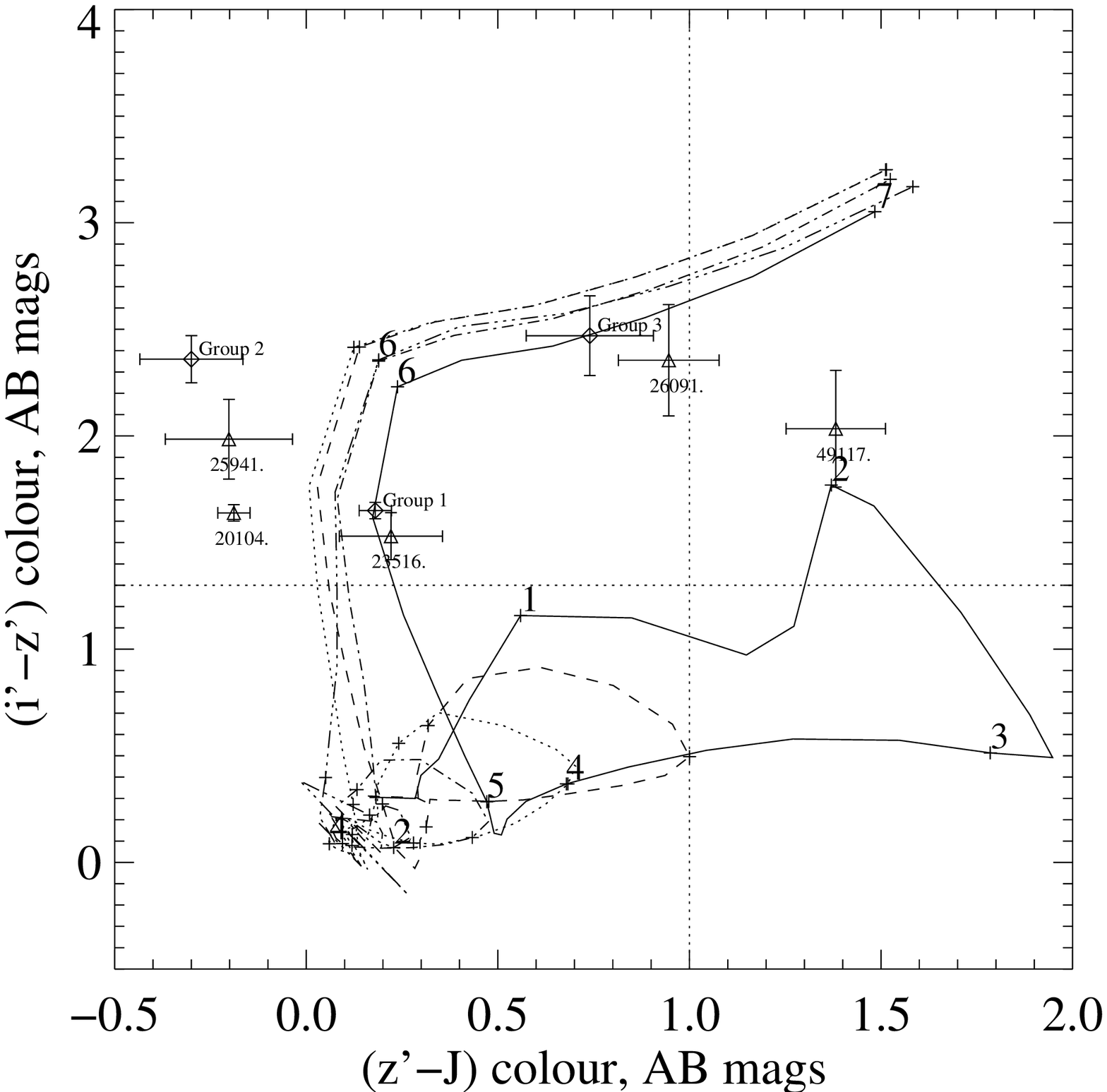}}
\resizebox{0.48\textwidth}{!}{\includegraphics{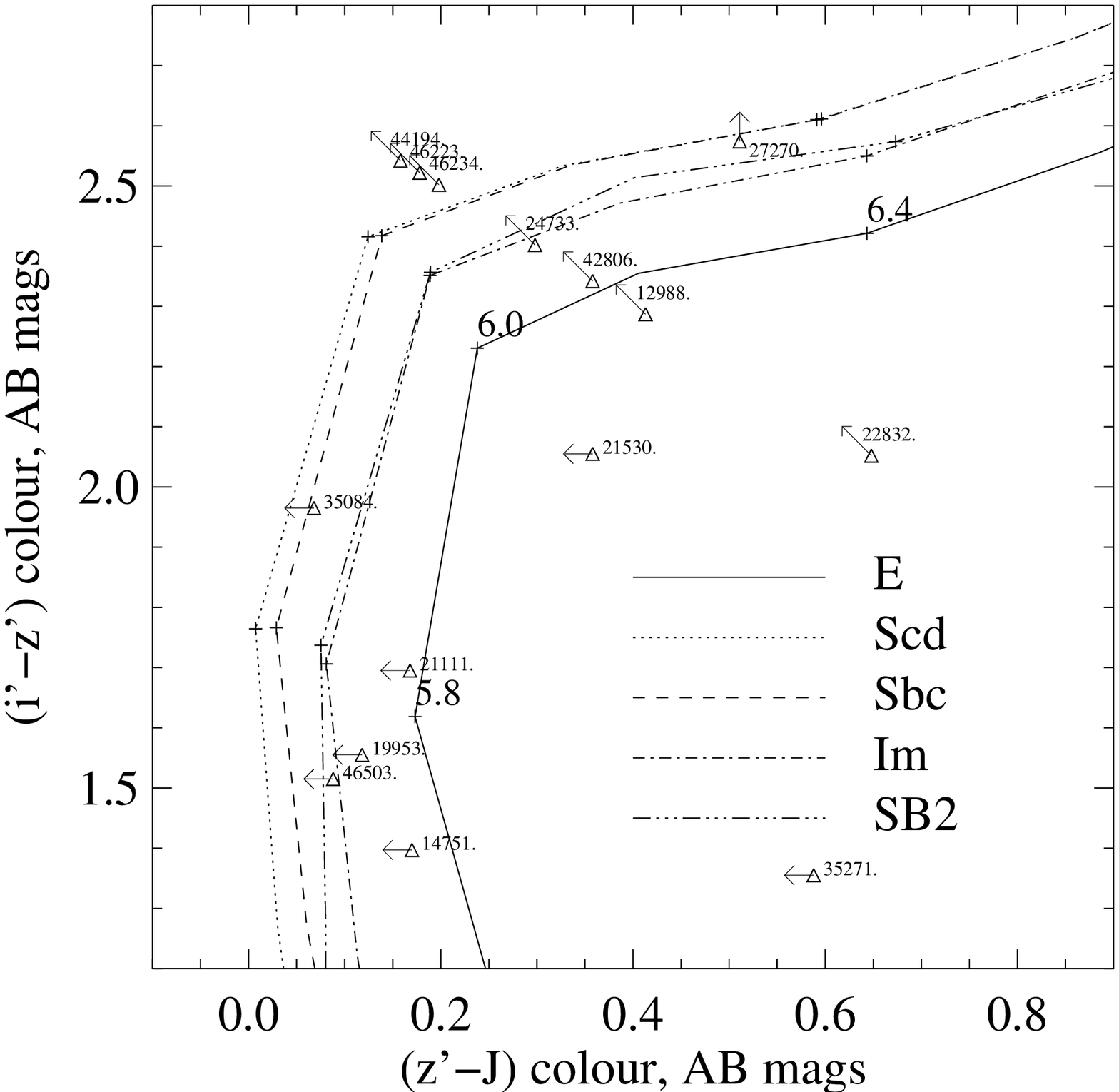}}
\caption{$z'-J$ v $i'-z'$ Colour-colour diagrams.  The evolutionary
  tracks followed by a starburst spectrum \citep{ki96} and the
  \citet{co80} empirical galaxy templates are plotted as lines (as in
  legend of of the right hand figure).  The colours of the $z\simeq6$
  candidate galaxies of Bunker et al.\ (2004) are indicated with solid
  symbols.  Colours of the galaxy groups as described in the text are
  indicated by diamonds.
  Objects for which only 3$\sigma$ limits
  are available are shown separately on the right hand figure for
  clarity.}
\label{fig:izj}
\end{figure*}

\begin{figure}
\resizebox{0.48\textwidth}{!}{\includegraphics{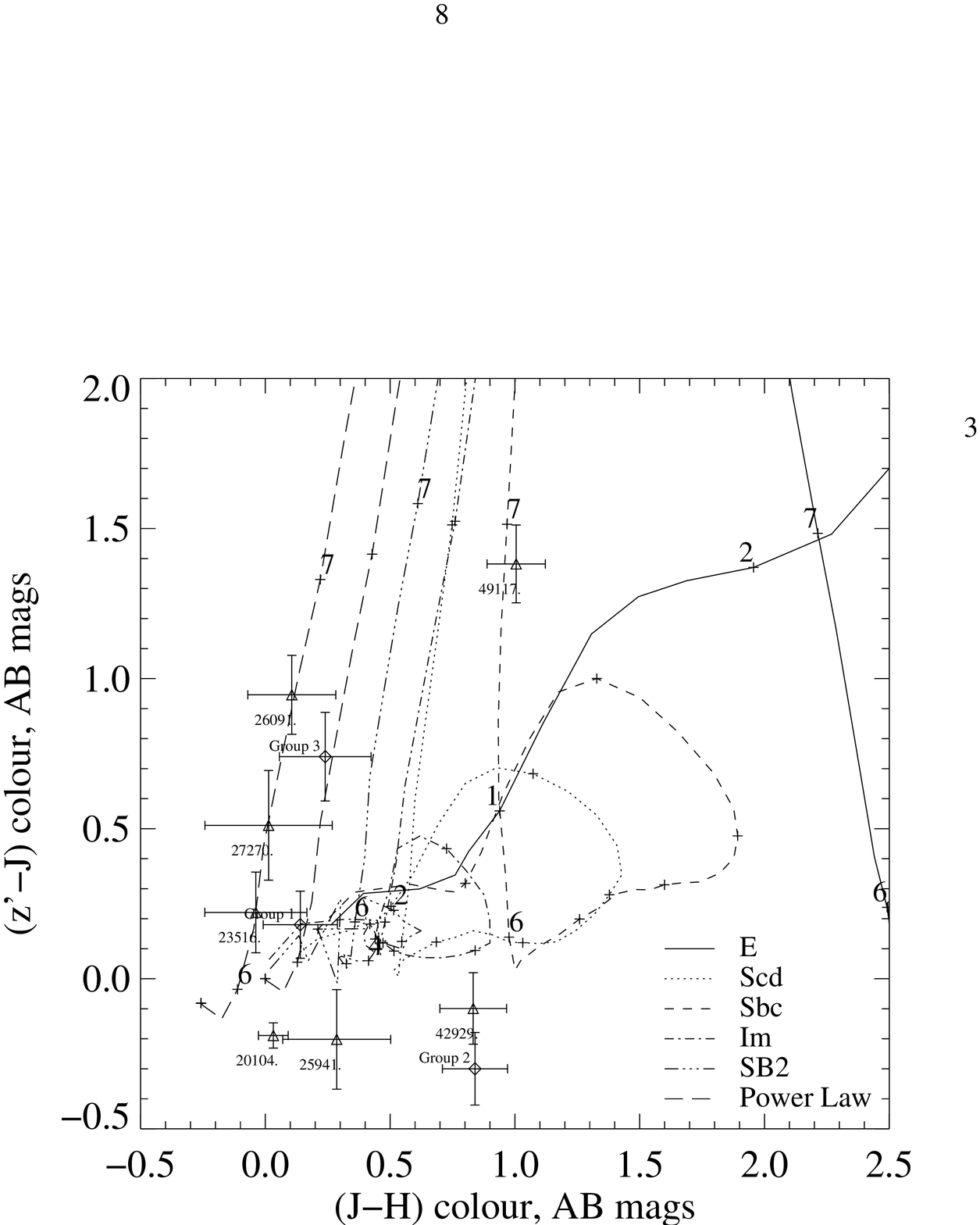}}
\caption{$J-H$ v $z'-J$ Colour-colour diagrams.  The evolutionary
  tracks followed by a \citet{ki96} starburst spectrum and the
  \citet{co80} empirical galaxy templates are plotted as lines (as in
  figure \ref{fig:izj}). In addition tracks for a synthetic spectra
  with power law slopes of $\beta=-2.0$ and $-2.5$
  ($f_\lambda=\lambda^\beta$) are shown (long dashes).  The colours of
  the $z\simeq6$ candidate galaxies of \citet{b04} are indicated with
  solid symbols. The colours of $i'$-drop groups are
  indicated with diamond symbols.}
\label{fig:zjh2}
\end{figure}


\subsection{Near-IR properties of fainter I-drops in the HUDF}
\label{sec:stack}

\begin{figure}
\resizebox{0.48\textwidth}{!}{\includegraphics{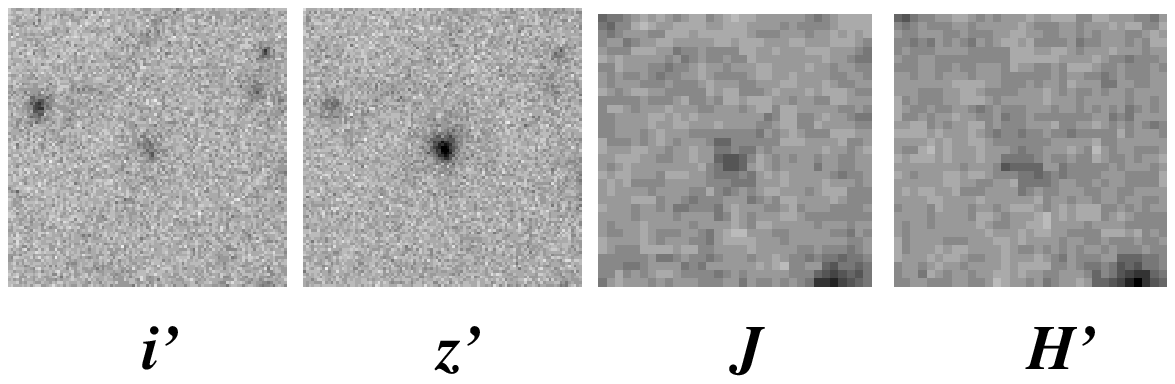}}
\caption{Stacked images of the nine objects undetected in the NICMOS
$J$ and $H$ band images which are more than 1\arcsec\ from their
nearest neighbour. The combined $i'$-drop sample is well detected in
both near-IR wavebands. The regions shown are 3\arcsec\ by 3\arcsec\
in size }
\label{fig:stack}
\end{figure}

Fourteen candidate high redshift galaxies identified by Bunker et al.\
(2004) lie within the NICMOS field yet are undetected to $3\sigma$ in the 
$J$ and $H$-band images. Of these, nine are isolated, lying
more than an arcsecond away from their nearest neighbours. In order to
investigate the properties of this fainter sub-catalogue of $i'$-drop
objects, sections of the NICMOS images, centered on the $z'$-band
coordinates of the 9 isolated candidate objects, were 
stacked. The resultant images (shown in figure \ref{fig:stack}) were then 
analysed with the SExtractor software
as described in section \ref{sec:cat}.

The background noise in the stacked images is a factor of $\sqrt{9}$
lower than in the original NICMOS images, allowing the average
properties of these candidates to be probed to a depth over a
magnitude fainter in each band. Their mean photometry is shown in
table \ref{tab:phot}. Clearly, this analysis assumes that the 9
$i'$-drop $z\approx6$ candidates considered are similar in colour and
that the mean colour is not heavily skewed by a single source.
However, the objects analysed are similar in $z'$ flux (spanning less
than 1 magnitude) and have near-IR limits in the NICMOS images which
rule out identification as low redshift elliptical galaxies.
The mean colours of these faint galaxies are bluer than
those for which secure NICMOS detections are available, suggesting a steep
mean rest-frame UV spectral slope (see section \ref{sec:zbeta}).

\section{Discussion}
\label{sec:disc}

\subsection{Faint Contaminants in $\mathrm{i'}$-Drop Selection}
\label{sec:cont}

The very red, old stellar populations of elliptical galaxies at around
$z=1-2$ (Extremely Red Objects, or EROs) allow them to contaminate any
sample of objects selected using a simple $i'-z'$ colour criterion (as
illustrated by figure \ref{fig:izj}, see also figure 2 of Stanway,
Bunker \& McMahon 2003).  The degree of sample contamination has long
been uncertain, affecting the analysis of population statistics at
high redshift.

Utilising the v0.5 release of data from the deep Great Observatories
Origins Deep Survey (GOODS), we have previously defined a sample of 20
$i'$-drop objects with $z'<25.6$ in the 300 arcmin$^2$ GOODS fields
\citep{st03,st04a}.  Of these 2 (10\%) are probable EROs and a further
six (30\%) cool Galactic stars giving a contamination at the bright
end of 40\%.  \citet{di04}, working with somewhat deeper GOODS data
and to a lower signal-to-noise limit reported a contamination fraction
of their $i'$-drop sample that increased at fainter magnitudes, reaching
as high as $\sim$45\% for resolved galaxies (c.f. 10\% for our
bright sample), although they report that their catalogue was affected
by a number of spurious sources not expected to be a problem in the
HUDF. \citet{bo04}, also using GOODS (v1.0) and with a similar
selection, reported a contamination of their $i'$-drop sample by lower
redshift ellipticals of 11\% based on the presence of flux in the $B$
or $V$ bands (unlikely for galaxies at $z\sim6$).

The availability of NICMOS imaging in the central portion of the HUDF
field allows lower redshift galaxies to be cleanly excluded from our
sample, as illustrated in figures \ref{fig:zjh2}, and \ref{fig:izj}.
Of the 27 $i'$-drop objects for which NICMOS infrared data is
available, only one (49117) has $z'-J$ colours which may be consistent
with those of $z\simeq2$ elliptical galaxies.

\begin{figure*}
\resizebox{0.7\textwidth}{!}{\includegraphics{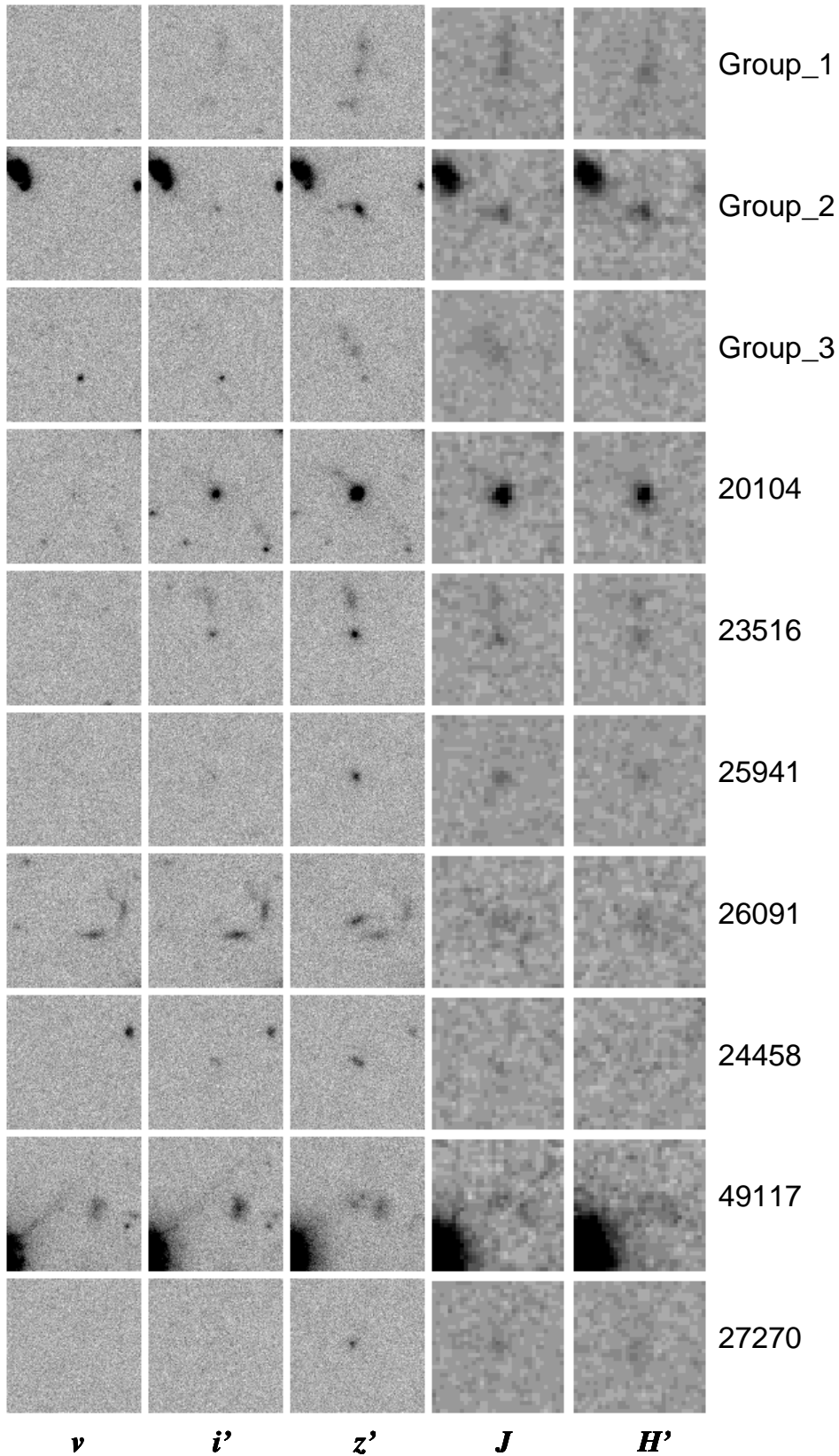}}
\caption{Postage stamp images from the ACS and NICMOS
for the first 10 objects listed in
Table \ref{tab:phot}. 
The regions shown are 3\arcsec\ by 3\arcsec\ in size and
orientated with north upwards. As can be seen, object 49117 is
contaminated in several bands by flux from its near neighbours and a
diffraction spike, while the groups of objects 
24123/24228/23972, 42929/42806 and 44154/44194 are each blended into one source in
the near-IR bands (labelled 1, 2 and 3 respectively in table
\ref{tab:phot}).}
\label{fig:stamps}
\end{figure*}

Object 49117 is likely tp be subject to contamination from two
neighbouring objects: one star, bright in the $J$ band ($J_{AB}=22$)
and separated from the candidate by 2\farcs3, and a second galaxy of
comparable $J$ magnitude to the candidate and separated from it by
only 0\farcs6 ($<2\times FWHM$, see figure \ref{fig:stamps}),
unresolved from it. None of the objects undetected in $J$ and $H$ have
colour limits consistent with lying at low redshift. This allows us to
place an upper limit on the low redshift galaxy contamination in this
$i'$-drop selected catalog of at most 1 objects in 27 ($<4\%$), or more
likely no $z\simeq2$ elliptical galaxies in this sample.

All these objects are also resolved in the $z'$-band, the only
unresolved $i'$-drop object reported in this field by \citet{b04}
having been excluded for having significant $V$-band flux. When
combined with the near-IR evidence from the NICMOS UDF exposure, this
suggests that the contamination of this $i'$-drop sample is less than
$5$\% - significantly lower than has been found by previous $i'$-drop
studies at brighter magnitudes, suggesting that we may be seeing
through the Galactic disk at these faint magnitudes.

\subsection{Redshift Discrimination}
\label{sec:zphot}

The $i'$-drop method is, in principle, sensitive to any galaxy bright
in the rest-frame ultraviolet and lying at redshifts between $z=5.6$
and $z=7$ (where the Lyman-break shifts out of the $z'$ filter).
However, the sharp fall off in filter transmission beyond 8500\AA\ 
biases the selection towards the low redshift end and renders unlikely
the detection of galaxies beyond $z\approx6.5$.  



The availability of extremely deep imaging across a wide wavelength
range, allows the photometric redshift method to be used
with good results. The publically available software package
\textit{hyperz} \citep{bo00} was used to calculate the photometric
redshifts of each object in table \ref{tab:phot} given their
magnitude, or magnitude limit, in the $v$, $i'$, $z'$, $J$ and $H$
bands (spanning $\lambda_{\mathrm rest}=4000-18000$\AA). This package
performs $\chi^2$ fitting of the object flux to a set of predefined
template spectra, selecting both the best fitting template and the
best fitting redshift simultaneously.  \citet{co80} E, Im, and Sbc
templates were used, together with power law spectra of the form
$f_\lambda\propto\lambda^\beta$ with $\beta=$-1.0, -1.5, -2.0, -2.2,
-2.5 (where $\beta=-2.2$ is suggested by the infrared colours of the
$i'$-drop sample, see section \ref{sec:zbeta}).

\begin{figure}
\resizebox{0.48\textwidth}{!}{\includegraphics{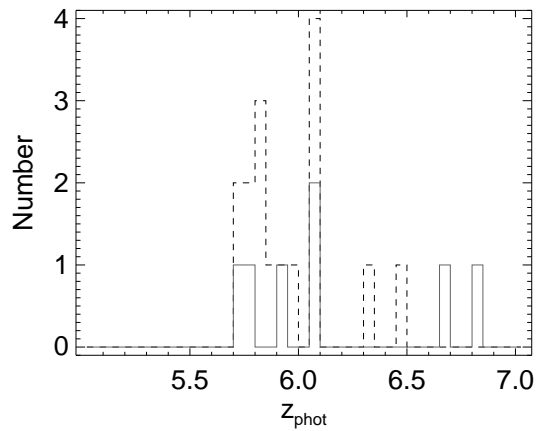}}
\caption{Photometric redshifts for the $i'$-drops in table
\ref{tab:phot}. 
The dotted lines are for the sample without neighbours and the
dashed line is for objects with photometry that may 
be effected by neighbours.
These redshifts were derived using the $hyperz$
template fitting package and a mixture of \citet{co80} templates and
power law spectra with slopes in the range $1.0<\beta<2.5$.} 
\label{fig:zphot}
\end{figure}

Trial redshifts were varied in the range $0<z_\mathrm{phot}<7$ at
intervals of $\Delta z=0.05$.  All sources were found to lie above
$z_\mathrm{phot}>5.6$ using this method.  To improve redshift
discrimination the range was further narrowed to $5<z_\mathrm{phot}<7$
for the remaining objects, with a redshift interval $\Delta z=0.02$,
resulting in the photometric redshifts presented in table
\ref{tab:phot} and illustrated in figure \ref{fig:zphot}. All objects
were best fit by power law spectra with slopes $\beta\ge-2.0$ with the
exception of objects 49117 (CWW E).

One $i'$-drop in this field has a confirmed spectroscopic redshift,
object 20104 at $z_\mathrm{spec}=5.83$ \citep{di04,st04a,st04b}. Its
photometric redshift is consistent with this
($z_\mathrm{phot}=5.82\pm0.01$).  
The mean $z_\mathrm{phot}$
of the total $i'$-drop sample is $6.02\pm0.30$. 
 The mean $z_\mathrm{phot}$ of the $i'$-drop sample using only the
objects(16) that have photometry unaffected by blending is
$5.96\pm0.22$ and $6.16\pm0.43$ for those that may be effected by
blending with companion objects(7).  
Hence a central assumed redshift of
6.0 \citep{b04} is in good agreement with the data.

\subsection{Effect of Spectral Slope and Line Contamination}
\label{sec:zbeta}

Interestingly, those objects for which measured colours are available
indicate an average spectral slope steeper than $\beta=-2.0$,
as do the mean colours of fainter objects detected in the stacked
image. The evidence for this trend is clear in figure
\ref{fig:zjh2} as the $J-H$ colour has a longer baseline than $z'-J$
and hence is more sensitive to spectral slope.  Although the error on
each colour is large, the colours of most of those objects for which
photometry is available lie between the redshift tracks predicted for
objects with spectral slopes of $\beta=-2.5$ and $\beta=-2.0$.

Strong line emission in the $J$-band might explain these observed
colours.  Narrow-band selected Lyman-$\alpha$ emitters at $z\sim5.7$
are known to have strong emission lines, with typical rest-frame
equivalent widths (EWs) of 50-150\AA\ \citep{aj03,hu04} and in some
cases much higher ($W_\mathrm{rest}>200$\AA\ e.g. Malhotra \& Rhoads
2002, Rhoads et al.\ 2003). In the observed frame at $z\sim6$, these
correspond to equivalent widths of $\sim$1000\AA. The $F110W (J)$-band
filter on NICMOS is broad with a FWHM of 5880\AA. Hence a line emitter
with an observed equivalent width of 1000\AA\ could contribute up to
17\% of the flux in this band or 0.2 mag. This, in fact,
underestimates the potential contribution of line flux to the $J$-band
magnitude since line blanketing below the Lyman break will suppress
the $J$-band flux for objects at redshifts greater than 5.6.

The observed colours are thus consistent with a redder intrinsic
rest-frame UV spectral slope (due to stellar continuum emission) of
$\beta=-2.0$, if every object has a powerful emission line. The single
$i'$-drop object in this field with a spectroscopically confirmed
emission line, however, has a much lower rest-EW
($W_\mathrm{rest}\sim30$\AA, contributing $<$5\% of the $z'$ flux,
Stanway et al.\ 2004a, Dickinson et al.\ 2004), as do several other
$i'$-drop objects which have now been confirmed spectroscopically
\citep{b03,st04b}.

 
An alternative explanation for the observed near-IR colours of this
sample may be evolution in the intrinsic spectral slope of starbursts
from that observed at lower redshifts. The observed rest-frame UV
spectral slope in star forming galaxies is influenced by two factors:
the intrinsic spectral energy distribution of the stellar population
(shaped by the initial mass function (IMF), and star formation
history) and the reddening due to dust extinction. The Lyman break
population at $z=3$ has a typical observed spectral slope of
$\beta=-1.5\pm0.4$ \citep{ad00}, as expected for dust-reddened
starburst galaxies with an intrinsic spectral slope of $\beta=-2.0$.
\citet{le99}, using their Starburst99 stellar synthesis code, find
that starbursts less than $\sim$ 200 Myr in age are well approximated
by the power law ($f_\lambda\propto\lambda^\beta$) model.  They
predict an intrinsic spectral slope at 1500\AA\ of $\beta=-2.5$ for
young (1-10 Myr), low metallicity starbursts (Leitherer et al.\ 1999,
figures 71 \& 72), independent of evolution and initial mass function
(IMF) effects (where IMFs considered include the classical
\citet{sa55} IMF, a Salpeter IMF truncated at 30M$_\odot$ and a
steeper IMF with an enhanced proportion of low mass stars). In these
young systems, hot OB type stars dominate the rest frame ultraviolet
continuum. The predicted $\beta$ falls to -2.0 on an IMF dependent
timescale of the order of $100$ Myr and falls more slowly with
increasing metallicity.  

These models do not account for the contribution to the UV spectral
slope due to dust reddening. The typical colour excess from the
stellar continuum of Lyman break galaxies at $z\sim3$ \citep{st99} is
E$_S$($B-V$)=0.16.  \citet{ca00}, studying a large sample of local
starburst galaxies, found a relation between colour excess, E($B-V$)
and observed spectral slope, $\beta$, of $E(B-V)=0.52(\beta-\beta_0)$
where $\beta_0$ is the intrinsic spectral slope, although the scatter
on this relation is considerable (approx 0.5 mag). If the dust
properties of the very young starbursts at $z\simeq6$ are similar to
those of Lyman break galaxies at lower redshift, this would imply an
intrinsic spectral slope of -2.9 for objects with $\beta=-2.2$
(typical for our sample, see sections \ref{sec:zphot} and
\ref{sec:zbeta}), outside the range of slopes predicted from starburst
models. This suggests that, if the $J-H$ colours -- and photometric
redshifts -- are good indicators of spectral slope, this $i'$-drop
sample is less affected by dust reddening than Lyman break galaxies at
lower redshift.

Given the fraction of $i'$-drops that are observed to be in close
pairs which may be interacting systems, these near-IR observations
suggest that we may be observing young starbursts triggered by an
epoch of merger activity.


\section{Implications for Star Formation History}
\label{sec:sfh}

The linear relation between star formation rate (SFR) and UV luminosity
density at $\lambda_\mathrm{rest}=1500$\AA\ \citep{ma98} has been
widely used to study the volume-averaged star formation history of the
universe (e.g. Madau et al.\ 1996, Steidel et al.\ 1999). The
broad-band photometry of a source may be used to determine its
luminosity density, if the redshift and appropriate $k$-correction
(from the effective wavelength of the filter to 1500\AA) is known.
Clearly the $k$-correction for UV luminous galaxies depend on their
spectral slope, $\beta$.

Galaxies selected by the $i'$-drop technique present several difficulties in
this regard. Firstly, with no strong constraint from bands at longer
wavelength than $z'$ ($\lambda_\mathrm{eff}$=9200\AA,
$\lambda_\mathrm{rest}=1300$\AA\ at $z=6$), the spectral slope of
these objects has been unknown and a value of $\beta=-2.0$ has been
assumed (e.g. Stanway, Bunker \& McMahon 2003) or a lower redshift
Lyman-break galaxy has been adopted as a template (e.g. Giavalisco et
al.\ 2004).  Furthermore, since the redshift distribution of objects is much
broader than that of galaxies selected using the more traditional
two-colour Lyman-break approach, luminosity densities have
been calculated by placing the candidate objects at some
model-dependent redshift, representative of the sample
as a whole. For $i'$-drop samples this representative redshift has
varied between $z=6.0$ \citep{b04} and $z=5.74$ \citep{gi04}
depending on details of models and the selection criteria.

Clearly, if the majority of objects lie at redshifts much higher or
much lower than has been estimated, the star formation density
calculated in earlier works will be unreliable. While the near-IR
data now available for $i'$-drop candidates in the NICMOS UDF do not
place a tight constraint on the redshift of individual objects,
their properties (discussed in sections \ref{sec:zphot} and
\ref{sec:zbeta}) are reassuringly consistent with those predicted in
simulations for the redshift range $z=5.6-6.5$ \citep{st04b,gi04,
  bo04}, with a median colour consistent with lying at $z\simeq6$ as
expected. However, as discussed in section \ref{sec:zbeta}, there is some
evidence for a slightly steeper spectral slope than the assumed value
of $\beta=-2.0$. A slope of $\beta<-2.0$ would imply that the
$k$-corrections hitherto applied to the data have led to a slight
overestimation of the star-formation density.

\begin{figure}
\resizebox{0.48\textwidth}{!}{\includegraphics{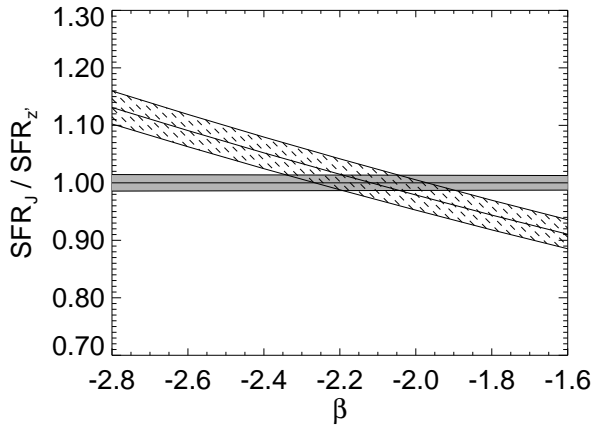}}
\caption{The ratio of SFRs derived from $J$ and $z'$ band fluxes 
  for those objects with secure $J$ band fluxes (objects 20104, 25941,
  27270 \& groups 1 and 3) for different spectral slopes.  Solid
  regions indicate the $1\sigma$ uncertainty on star formation rates
  derived from $z'$ magnitudes ($\lambda_\mathrm{rest}=1400$\AA),
  line-filled regions indicate star formation rates from the $J$-band
  (observed at $\lambda_\mathrm{rest}=1600$\AA). In each case the
  1500\AA\ luminosity density has been corrected using the spectral
  slope indicated and assuming the objects lie at $z=6.0$.}
\label{sfrcomp}
\end{figure}

Figure \ref{sfrcomp} compares the mean SFRs derived from the $z'$ and
$J$ band filters, centered at $\lambda_\mathrm{rest}=1400$\AA\ and
$1600$\AA\ at $z=6.0$ respectively, for the subsample of objects
isolated from lower redshift sources and for which a
secure $J$ band detection was made.  The luminosity-SFR relation of
\citet{ma98} (which uses a \citet{sa55} initial mass function) is
used ($L_{UV} = 8\times10^{27}$ (SFR/M$_\odot$ yr$^{-1}$) ergs
s$^{-1}$ Hz$^{-1}$).  Objects were assumed to lie at the mean redshift
of $z=6.0$ \citep[see][]{b04}.
The total flux of these seven objects in each waveband was
summed before conversion to star formation rates
using the spectral slope specified.  If the value of $\beta$ used to
calculate the $k$-correction accurately represents the population mean
slope, then the star formation rates derived from any two bands
should converge.

The photometric errors at these faint magnitudes, and
the consequent uncertainties in the SFRs, are
large.  Nonetheless, the slope which gives the best agreement
between $z'$ and $J$-band derived star formation rates in the
$i'$-drop sample ($\beta=-2.12\pm0.15$) does suggest a slightly
steeper slope than $\beta=-2.0$, as expected given the
$J-H$ colour distribution. 
In practise, varying the spectral slope between $\beta=-2.0$ and
$\beta=-2.2$ changes the calculated star formation rate by only
$\approx6$\%. The derived value of $\beta$ is also consistent with a
slope flat in $f_\nu$ ($\beta=-2.0$) used in earlier computations of
star formation rates (e.g. Bunker et al.\ 2004).

%

The infrared properties of this $i'$-drop sample do suggest further
caveats to be applied when calculating the star formation density at
$z\simeq6$. If, $i'$-band drops are indeed young starbursts, then the
\citet{ma98} luminosity-star formation rate relation may be
underestimating the star formation density. Stellar populations $<$10
Myr old yield SFRs more than 50\% higher than given by the
\citet{ma98} calibration when modelled using stellar synthesis
techniques \citep{le95,ke98,le99}.

On the other hand, a top-heavy IMF rich in massive stars has been
hypothesised at very high redshifts (e.g. $z>10-15$, Clarke \& Bromm
2003) and such an IMF could explain the steep UV spectral slope
observed.  If so, this would also alter the luminosity-SFR calibration
which extrapolates the formation of high mass stars to the low mass
regime. A top heavy IMF would reduce the star formation rate inferred
from a given 1500\AA\ luminosity.

\section{Conclusions}
\label{sec:conc}

\begin{enumerate}
\item The near-infrared properties of a faint sample of $i'$-drop
objects (defined by Bunker et al.\ 2004) has been investigated, using
the deep $F110W$ and $F160W$ images of the HST/ACS Ultra Deep Field.

\item The infrared colours of all objects not suffering from
contamination by near neighbours are consistent with a high redshift
interpretation. This suggests that the contamination of this $i'$-drop
sample is smaller than that observed at brighter magnitudes.

\item The spectral slopes inferred from near infrared colours are
consistent with $\beta=-2.0$ although there is evidence for a marginally 
steeper spectral slope of $\beta=-2.2\pm0.2$.

\item These steep spectral slopes suggest that the dust extinction of
the sample is small and that the $i'$-drop population may comprise
galaxies with young starbursts - possibly triggered by mergers
(several close systems of galaxies are observed in this sample).

\item The low levels of contamination, and the steep spectral slope,
inferred from near-infrared data supports the validity of our previous
star formation density estimates based on $i'$-drop samples.
\end{enumerate}

\subsection*{Acknowledgements}

Based on observations made with the NASA/ESA Hubble Space Telescope,
obtained from the Data Archive at the Space Telescope Science
Institute, which is operated by the Association of Universities for
Research in Astronomy, Inc., under NASA contract NAS 5-26555. These
observations are associated with programs \#9803 and \#9978. We thank
Paul Hewett, Richard Ellis and Malcolm Bremer for useful discussions.
ERS acknowledges a Particle Physics and Astronomy Research Council
(PPARC) studentship supporting this study.  We also thank Steve
Beckwith and the HUDF team for making these high quality images
publically available.

\bsp


\begin{thebibliography}{}

\bibitem[\protect\citeauthoryear{Adelberger \& Steidel}{2000}]{ad00}
Adelberger, K. L., Steidel, C. C., 2000, ApJ, 544, 218 

\bibitem[\protect\citeauthoryear{Ajiki et al.\ }{2003}]{aj03}
Ajiki et al., 2003, AJ, 126, 2091

\bibitem[\protect\citeauthoryear{Beckwith, Somerville \& Stiavelli } {2003}]{be03}
Beckwith, S., Somerville, R., Stiavelli, M. 2003, 
STScI Newsletter vol 20 issue 04

\bibitem[\protect\citeauthoryear{Bertin \& Arnouts}{1996}]{ba96} 
Bertin E., Arnouts S., 1996, A\&AS, 117, 393

\bibitem[\protect\citeauthoryear{Bolzonella, Miralles \& Pell\'o}{2000}]{bo00}
Bolzonella, M., Miralles, J.~M., Pell\'o, R., 2000, A\&A, 363, 476

\bibitem[\protect\citeauthoryear{Bouwens et al.\ }{2003a}]{bo03a}
Bouwens R. et al., 2003a, ApJ, 593, 640

\bibitem[\protect\citeauthoryear{Bouwens et al.\ }{2003b}]{bb03}
Bouwens R. et al., 2003b, ApJ, 595, 589

\bibitem[\protect\citeauthoryear{Bouwens et al.\ }{2004}]{bo04}
Bouwens R. et al., 2004, preprint (astro-ph/0403167)

\bibitem[\protect\citeauthoryear{Bremer et al.\ }{2004}]{br04}
Bremer, M. N., Lehnert, M. D., Waddington, I., Hardcastle, M. J., Boyce, P. J., Phillipps, S., 2004, MNRAS, 347, L7

\bibitem[\protect\citeauthoryear{Bunker et al. }{2003}]{b03}
Bunker, A. J., Stanway, E. R., Ellis, R. S., McMahon, R. G., McCarthy, P. J.,
 2003, MNRAS, 342, L47 

\bibitem[\protect\citeauthoryear{Bunker et al. }{2004}]{b04}
Bunker, A. J., Stanway, E. R., Ellis, R. S., McMahon, R. G., 
2004, preprint (astro-ph/0403223)

\bibitem[\protect\citeauthoryear{Clarke \& Bromm}{2003}]{cl03}
Clarke, C. J., Bromm, V., 2003, MNRAS, 343, 1224

\bibitem[\protect\citeauthoryear{Calzetti et al.\ }{2000}]{ca00}
Calzetti, D., Armus, L., Bohlin, R. C., Kinney, A. L., Koornneef, J.,
Storchi-Bergmann, T., 2000, ApJ, 533, 682

\bibitem[\protect\citeauthoryear{Coleman, Wu \& Weedman }{1980}]{co80}
Coleman G.~D., Wu C.-C., Weedman D.~W., 1980, ApJS, 43, 393

\bibitem[\protect\citeauthoryear{Connolly et al.\ }{1997}]{co97}
Connolly A. J., Szalay A. S., Dickinson M., Subbarao M. U., Brunner R. J., 1997, ApJ, 286, L11

\bibitem[\protect\citeauthoryear{Dickinson et al.\ }{2004}]{di04}
Dickinson M. et al., 2004, ApJ, 600, L99

\bibitem[\protect\citeauthoryear{Fan et al.\ }{2003}]{fa03}
Fan X. et al., 2003, AJ, 125, 1649


\bibitem[\protect\citeauthoryear{Fern\'andez-Soto, Lanzetta \& Yahil }{1999}]{fe99}
Fern\'andez-Soto A., Lanzetta K. M., Yahil A., 1999, ApJ, 513, 34

\bibitem[\protect\citeauthoryear{Fujita et al.}{2003}]{fu03}
Fujita et al., 2003, AJ, 125,13

\bibitem[\protect\citeauthoryear{Giavalisco et al.\ }{2004}]{gi04}
Giavalisco M. et al., 2004, ApJ, 600, L103

\bibitem[\protect\citeauthoryear{Guhathakurta, Tyson \& Majewski }{1990}]{guh90}
Guhathakurta, P., Tyson, J. A., Majewski, S. R.,, 1990, ApJL, 357, 9

\bibitem[\protect\citeauthoryear{Hu \& McMahon}{1996}]{hm96}
Hu, E.~M., McMahon, R.~G., 1996, Nature, 382, 281

\bibitem[\protect\citeauthoryear{Hu, McMahon \& Cowie }{1999}]{hu99}
Hu E.~M., McMahon R.~G., Cowie L.~L., 1999, ApJ, 522, L9

\bibitem[\protect\citeauthoryear{Hu et al.\ }{2002}]{hu02}
Hu, E.M. et al., 2002, ApJL, 568, 75

\bibitem[\protect\citeauthoryear{Hu et al.\ }{2004}]{hu04}
Hu, E. M., Cowie, L. L., Capak, P., McMahon, R. G., Hayashino, T., Komiyama, Y., 2004, AJ, 127,563

\bibitem[\protect\citeauthoryear{Kennicutt }{1998}]{ke98}
Kennicutt, R. C., 1998, ARA\&A, 36, 189

\bibitem[\protect\citeauthoryear{Kinney et al.\ }{1996}]{ki96}
Kinney, A. L., Calzetti, D., Bohlin, R. C., McQuade, K.,
Storchi-Bergmann, T., Schmitt, H. R.,
1996, ApJ, 467, 38

\bibitem[\protect\citeauthoryear{Kodaira et al.\ }{2003}]{ko03}
Kodaira, K. et al., 2003, PASJ, 55, L17

\bibitem[\protect\citeauthoryear{Koornneef }{1983}]{ko83}
Koornneef J., 1983, A\& A, 128, 84

\bibitem[\protect\citeauthoryear{Lanzetta et al.\ }{2002}]{lan02}
Lanzetta K.~M., Yahata N., Pascarelle S., Chen H., Fern{\' a}ndez-Soto A., 2002, ApJ, 570, 492

\bibitem[\protect\citeauthoryear{Lanzetta, Yahil \& Fernandez-Soto }{1996}]{lan96}
Lanzetta K. M., Yahil A., Fernandez-Soto A., 1996, Nature, 381,759

\bibitem[\protect\citeauthoryear{Leitherer et al.\ }{1999}]{le99}
Leitherer et al., 1999, ApJS, 123, 3

\bibitem[\protect\citeauthoryear{Leitherer \& Heckman }{1995}]{le95}
Leitherer, C., Heckman, T. M., 1995, ApJS, 96, 9

\bibitem[\protect\citeauthoryear{Madau, Pozzetti \& Dickinson } {1998}]{ma98}
Madau P., Pozzetti L., Dickinson M., 1998, ApJ, 498, 106

\bibitem[\protect\citeauthoryear{Madau et al.\ }{1996}]{ma96}
Madau P., Ferguson H.~C., Dickinson M.~E., Giavalisco M., Steidel C.~C.,
Fruchter A., 1996, MNRAS, 283, 1388

\bibitem[\protect\citeauthoryear{Madau}{1995}]{ma95}
Madau P.,  1995, ApJ, 441, 18

\bibitem[\protect\citeauthoryear{Oke \& Gunn }{1983}]{og83}
Oke J.~B., Gunn J.~E., 1983, ApJ, 266, 713

\bibitem[\protect\citeauthoryear{Salpeter } {1955}]{sa55}
Salpeter E.~E., 1955, ApJ, 121, 161

\bibitem[\protect\citeauthoryear{Schlegel, Finkbeiner \& Davis }{1998}] {sfg98}
Schlegel D.~J., Finkbeiner D.~P., Davis M., 1998,
ApJ, 500, 525

\bibitem[\protect\citeauthoryear{Stanway, Bunker \& McMahon }{2003}]{st03}
Stanway E. R., Bunker A. J., McMahon R. G., 2003,
MNRAS, 342, 439

\bibitem[\protect\citeauthoryear{Stanway et al.\ }{2004a}]{st04a}
Stanway E. R. et al., 2004a, ApJ, 604, L13

\bibitem[\protect\citeauthoryear{Stanway et al.\ }{2004b}]{st04b}
Stanway E. R., Bunker, A. J., McMahon, R. G., Ellis, R. S., Treu, T., McCarthy, P. J., 2004b, ApJ, 607, 704

\bibitem[\protect\citeauthoryear{Steidel, Pettini \& Hamilton }{1995}]{st95}
Steidel C.~C. Pettini M., Hamilton D., 1995, AJ, 110, 2519

\bibitem[\protect\citeauthoryear{Steidel et al.\ } {1999}]{st99}
Steidel C.~C., Adelberger K.~L., Giavalisco M., Dickinson M.~E., 
Pettini M., 1999, ApJ, 519, 1

\bibitem[\protect\citeauthoryear{Steidel, Pettini \& Adelberger } {2001}]{st01}
Steidel C.~C., Pettini M., Adelberger K.~L., 2001, ApJ, 546, 665

\bibitem[\protect\citeauthoryear{Thompson et al.\ }{1999}]{th99}
Thompson R. I., Storrie-Lombardi, L. J., Weymann, R. J., Rieke,
M. J., Schneider, G., Stobie, E., Lytle, D., 1999, AJ, 117, 17.

\bibitem[\protect\citeauthoryear{Williams et al.\ }{1996}]{wi96}
Williams et al., 1996, AJ, 112, 1335

\bibitem[\protect\citeauthoryear{Yan, Windhorst \& Cohen } {2003}]{yan03}
Yan H. Windhorst R.~A., Cohen S., 2003, ApJ, 585, L93

\end{thebibliography}
\end{document}